\documentclass[letter,times,twocolumn,10pt]{article}
\pdfoutput=1
 \usepackage{usenix,epsfig,endnotes}

\usepackage[normalsize,bf]{caption}
\usepackage[compact]{titlesec}
\usepackage{times,amsmath,epsfig}

\usepackage{amsmath, amssymb, amsfonts}
\usepackage{epsfig, graphicx, subfigure}
\usepackage{url}
\usepackage{algorithmic}
\usepackage{wrapfig}
\usepackage{times}
\usepackage{algorithm}
\usepackage{latexsym}
\usepackage{mdwlist}

\def\etal{et al. }

\title{Where Have You Been? Secure Location Provenance for\\Mobile Devices}

\author{%
{Ragib Hasan, Randal Burns}%
\vspace{1.6mm}\\
\fontsize{10}{10}\selectfont\itshape
Department of Computer Science, Johns Hopkins University\\
3400 N. Charles Street, Baltimore, MD 21218, United States\\
\{ragib,randal\}@cs.jhu.edu\\
}

\date{}

\begin{document}
\maketitle
    \begin{abstract} 
With the advent of mobile computing, location-based services have recently gained popularity. Many applications use the location provenance of users, i.e., the chronological history of the users' location for purposes ranging from access control, authentication, information sharing, and evaluation of policies. However, location provenance is subject to tampering and collusion attacks by malicious users. In this paper, we examine the secure location provenance problem. We introduce a witness-endorsed scheme for generating collusion-resistant location proofs. We also describe two efficient and privacy-preserving schemes for protecting the integrity of the chronological order of location proofs. These schemes, based on hash chains and Bloom filters respectively, allow users to prove the order of any arbitrary subsequence of their location history to auditors. Finally, we present experimental results from our proof-of-concept implementation on the Android platform and show that our schemes are practical in today's mobile devices.
\end{abstract}
    \section{Introduction}
\label{sec:intro}

As mobile computing becomes popular, users are no longer physically confined to a fixed location. Location-based services have become very popular in recent years. In many scenarios, access control, authentication, and other important decisions can be made based on a user's current and past physical locations. The location history of a user is important in diverse applications such as supply-chain management, voter registration, offender tracking, customer-loyalty programs, reimbursement processing, perimeter security, and intrusion detection \cite{saroiu2009hotmobile}. However, when location history is used for any high-stakes application, we must ensure the trustworthiness of the location history. Real-life examples from the popular location-aware social game FourSquare show that, even when the incentive is to get free coffee or small discounts, people can often cheat by misreporting or manipulating their location history \cite{foursquare2}. 

The location provenance of a user is the history of the user's locations over time. To verify a user's claims about her location history, an auditor needs to verify her claims about individual locations and visit times, as well as the order of the her visits. Self-reported location information such as Global Positioning System (GPS)  traces or cell phone triangulation can be manipulated by malicious users to support false location claims. Continuous tracking and reporting of user's locations violate privacy, and is not scalable in distributed environments. A more feasible and scalable approach is to require the user to obtain proofs of presence from the locations she visits. To issue a proof, locations first ensure the user's presence in the area using secure localization techniques such as distance-bounding \cite{brands93eurocrypt,rasmussen2010distance}. Then, a proof can be issued to the user which can later be used to prove the user's presence and visit time to a third party auditor. 

Researchers have recently proposed schemes for generating location proofs that a user can present to an auditor in order to prove the user's presence at a particular location at a specific point in time \cite{gilbert2010hotmobile,saroiu2009hotmobile,waters2003secure}. But these schemes are susceptible to collusion attacks where the user and the location collude to create fake proofs. Existing location proof schemes also fail to provide proofs of  the user's chronological location history, since each location proof only attests visits to a single location. Verifying the order of a user's past locations is also complicated as different locations may not have a synchronized global clock, rendering local timestamps useless for finding the order of the location visits. Using a centralized timestamping service is not scalable, and also becomes a single point of failure. To make location provenance trustworthy, we must ensure the integrity of the chronological order of the location proofs and prevent collusion attacks that create false history. At the same time, we need to balance the tradeoff between the need to verify location history versus user privacy. A user should be able to prove any subsequence of her location provenance without revealing her entire location history. 

To illustrate the problem, we consider the following scenario:

\textit{Mallory owns a moving company that bills customers based on the distance traveled. In practice, most moving companies use a shared truck to move items. At any given time, the truck contains items belonging to several customers, who are billed separately. Alice and Bob both hire the moving company. After delivering the items, Mallory bills her customers based on a route she claims to have used. However, Alice and Bob want to verify the route claimed by Mallory before they would pay the bill. Mallory can present subsequences of her location provenance to Alice and Bob or their auditors for verification of the truck's presence in locations along the claimed routes. To protect the privacy of her customers, Mallory should reveal to each customer only the information pertaining to that customer's route. Alice and Bob also want to ensure that Mallory cannot create a false proof of visiting a location owned by her friend Colin.}

In this paper, we examine the problem of verifying a user's location history and provenance in untrusted distributed environments.  To address the shortcomings of existing location proof schemes, we introduce the notion of third party witnesses. To detect collusion between users and locations to create false location proofs, we require location proofs to be endorsed by third party witnesses. We also examine the problem of verifying the chronological order of location proofs. In particular, we focus on designing schemes that allow users to prove the chronological order of any arbitrary subsequence of their locations, without revealing their entire location history. Since location proofs are gathered by power and computationally constrained mobile devices, we also aim at designing efficient schemes.\\

\noindent\textbf{Contributions.~}
The contributions of this paper are as follows:
\begin{enumerate*}
\item We introduce a witness-endorsed collusion-resistant scheme for generation of location proofs.
\item We present a scheme for designing private location proofs that allows users to reveal their location history only at the desired granularity.
\item We design two privacy-preserving schemes -- one using hash chains, the other using Bloom filters -- for protecting the integrity of the chronological order of location proofs. Our schemes allow users to prove any arbitrary subsequences of their location provenance.
\item We evaluate the performance of our system on Android-based mobile phones. Our experiments show that our schemes for the creation and verification of location proofs and provenance are practical in today's mobile phones.
\end{enumerate*}

\noindent\textbf{Organization.~} The rest of the paper is organized as follows: Section \ref{sec:background} discusses applications of secure location provenance and the challenges in protecting location history. In Section \ref{sec:model}, we present our witness-endorsed location proof and provenance model, and discuss the system and threat models. Next, in Section \ref{sec:scheme}, we provide  details of the location provenance generation protocol and discuss our schemes for generating private location proofs and provenance. A security analysis of our scheme against different types of attacks is presented in Section \ref{sec:analysis}. We present a proof-of-concept practical implementation of the our schemes and experimental results in Section \ref{sec:evaluation}. Finally, we discuss related work in Section \ref{sec:related}, and conclude in Section \ref{sec:conclusion}.
	\section{Applications and Challenges}
\label{sec:background}

Provenance of an object provides the past history of the object, i.e., its origin, modifications, movements, and other events that happened throughout its lifetime \cite{simmhan05}. For example, provenance of a painting provides information about the painter, owners of the painting, information on exhibitions, etc. In computing, provenance has mostly been used for recording the history of data items \cite{buneman01icdt,simmhan05}. In this paper, we propose using provenance in another important application area -- the location history of mobile devices and users. In this section, we discuss potential real-life applications of secure location provenance. We also analyze different challenges that make location provenance difficult to protect against malicious adversaries.

\subsection{Applications}

Researchers have discussed many important applications of systems that uses location proofs. For example, Sariou \etal \cite{saroiu2009hotmobile} described six example applications: customer loyalty (i.e., verifying whether customers actually visited the store many times), environment-friendly incentives for transport (i.e., verifying that users followed an environment friendly route), fine-grained location-restricted content delivery, assisting in police investigation (i.e., verifying a suspect's alibi), voter registration (verifying that the voters actually live in the claimed constituency), and reducing auction frauds. We argue that in addition to location proofs for individual locations, the chronological order of these proofs is also important to verify the user's path. Location provenance of a user is important in situations when we make decisions based on the physical movement of the device. For example, if we know that a computing device was exposed to an untrustworthy operating environment, then we can deem the device to be insecure. Another real life example comes from secure verification of the supply chains of different objects. For example, consumers are often not willing to purchase meat or other food coming from an unknown place or an origin with questionable health standards.

The ability to check the location provenance, i.e. both the user's presence in individual locations and the path of the user, opens up new mobile application areas. As an example, location provenance can allow verification of employee reimbursement claims (based on the employees travel and location history). Location provenance can be used to implement new security policies, where only users with certain location histories can access a resource. Another example is airline security: a passenger must go through a specific route (terminal, security checkpoints, gate) before she can enter a plane. 

\subsection{Challenges}
Many issues complicate secure determination of a device's location over time. Here, we present some of the important issues and their impact on secure location verification.\\

\noindent\textbf{Trustworthiness of location information.~} If all users were trustworthy, then self-reported location claims would also be trustworthy. However, without attestation from other entities, we cannot use self-reported location claims from users as they may not be trustworthy. Location information reported by a location authority, i.e., an observer associated with a location, can also be untrustworthy if such locations are malicious or are compromised. So, we cannot depend on a single party's claims when building location proofs.

\noindent\textbf{Proof storage.~}
Where should the location proofs be stored? The proofs can be stored in a trusted central server. But in a large system, such a server will be a bottleneck. The proofs can also be stored locally at each of the locations, but it makes audits inefficient since the auditor would need access to these proofs and therefore contact each location in order to verify the user's claim. To make audits more efficient, it is better to store the proofs at the user's mobile device, which can build up a repository of proofs for locations the user visits over time. However, storing the proofs on the device also introduces security problems -- the user has full access to its storage and therefore can tamper with the proofs. So, any solution to this problem requires mechanisms that make such unauthorized modifications easy to detect.

\noindent\textbf{Collusion.~}
Many of the existing schemes for verification of location proofs assume that the locations that provide the location proof are trustworthy \cite{luo2010hotmobile,saroiu2009hotmobile,waters2003secure}. However, we argue that this assumption is flawed as in distributed environments, it is not realistic to consider location authorities from different security and administrative domains to be trustworthy. Users can collude with the location authorities to create false proofs of location. Such collusion can be motivated by monetary incentives. Such collusion can be real-time or post-facto. Users can also collude with location authorities to implicate an honest user. 

\noindent\textbf{Clock synchronization and timestamping.~}
To accurately determine the sequence of locations visited by a user, the auditor needs to check the order of location proofs. However, in a distributed environment, clocks of different verifiers/observers can be different. So, the timestamp provided by the proof issuer cannot be used to determine order. The timestamp from the clock of the user's mobile device is also not trustworthy, since it can be manipulated by the device owner to present a false timeline. A secure third party timestamping service can be used to timestamp a location proof once it is generated, but that scheme fails when the user and the location colludes. 

\noindent\textbf{Location granularity.~}
The granularity of location information can be a challenge when building a location proof. The position of a user can be described using different granularities.  The user may want to reveal information only in the appropriate granularity to different applications. For example, the user may be willing to reveal only coarse-grained location information to an untrusted or semi-trusted auditor. An auditor should not learn information in a finer granularity that what the user reveals.

\noindent\textbf{Privacy.~} 
Location privacy is an important issue when it comes to tracking the location provenance. We need to ensure the privacy of location information. While a user wants to reveal her location to the auditor, she should not be required to reveal all of her locations unless she wants to do so. An auditor should only be able to access location information that the user has authorized the auditor to do. To allow this, any scheme to provide location provenance should allow revealing any subset of location history to the auditor. At the same time, the auditor needs to be able to verify the order of the user's location when given any arbitrary subsequence of the user's location provenance. The next privacy issue arises from the privacy of users visiting a location. While the user might need endorsement from other users in finding its location, such witnesses should not be forced to reveal their identities. Witnesses should be able to anonymously endorse location information of other users. Finally, when a user asks for location information or an endorsement from a location, the location should not learn anything about the previous location of the user. 


\noindent\textbf{Identity and user presence.~} The identity of a user should be unforgeable and unique. Otherwise, users can create false identities and masquerade as their own witnesses. A user's identity can be bound to her public key or a digital certificate stored in her device.  A user identity should also be resilient, hard to remove from her mobile devices, and should not be easy to spoof or clone. A related issue is to ensure user's presence with the mobile device. To use a location proof based on the position of the user's mobile device,  we must ensure that the user is actually carrying the device. To provide a stronger binding between the user and the device, researchers have proposed different techniques such as biometrics \cite{saroiu2009hotmobile}.  For the rest of this paper, we will assume that the location of the user's mobile device indicates the users location. 

\noindent\textbf{Deployment.~}
Deployment of location proof systems must be incremental in nature. In the beginning, a location authority can cover a large area, say, a city, and provide coarse grained location proofs for the devices in the city. As more entities are willing to provide location proofs, more location proof authorities can be added, with finer grained location proofs.

	\section{A Model for Secure Location Provenance}
\label{sec:model}
In this section, we present a model for secure location provenance. We define secure location provenance as ensuring the integrity and privacy of the location history of a user. Many existing location proof schemes are subject to collusion attacks. In this section, we present a collusion-resistant, privacy-aware location provenance scheme. 

\subsection{Overview}
From time immemorial, witnesses have been used to attest to or endorse contracts between two parties that are untrustworthy or mutually suspicious. A witness is someone who is present when two other parties exchange information or agree on some common statement. We utilize the same idea to make location proofs robust and collusion resistant. Informally, a witness is another user or entity that is spatio-temporally co-located with the user at the same location. By requiring a witness to endorse location proofs, we reduce the threat of collusion between the user and the location to create false proofs. Below, we present the terminology used in the model:\\

\noindent\textbf{User.~} A \textit{User} is a mobile entity that moves from location to location. The user carries a mobile device with her which is used to detect her location and to store the location proofs. 

\noindent\textbf{Location.~} A \textit{Location} is a physical region with a finite area. A location is under the coverage of one or more location authorities. 

\noindent\textbf{Location authority.~} A \textit{Location authority} is a stationery entity that is responsible for providing location proofs for a particular location. 

\noindent\textbf{Location proof.~} A \textit{Location proof} is a proof which allows an auditor to verify that a user was physically present at a given location at a specific time.

\noindent\textbf{Endorsement.~} An \textit{Endorsement} of a location proof is a verifiable statement from a third-party supporting the location proof.

\noindent\textbf{Location provenance entry.~} A \textit{Location provenance entry} is a claim about a user's location at a certain time, supported by one or more location proofs and endorsements. It also includes ordering information which is used to determine the chronological order of location proofs. 

\noindent\textbf{Location provenance chain.~} A \textit{Location provenance chain} is a chronologically ordered sequence of location provenance entries.

\noindent\textbf{Witnesses.~} A \textit{Witness} is an entity who can endorse a location proof, attesting the presence of a user in a particular location belonging to an observer. Witnesses can themselves be mobile users. They can detect the presence of users within their reach.

\noindent\textbf{Auditor.~} An \textit{Auditor} is a semi-trusted entity that can verify the location history of a user. Given a subsequence of a user's location provenance chain, the auditor is able to verify whether the chain is valid or it has been tampered with, or it contains false statements.

\subsection{System Model}

We assume that users carry mobile devices capable of communicating with other devices and locations over Wi-Fi or Bluetooth. The devices have local storage for storing the location proofs. We assume that the device owner or user has full access to the storage and computation of the device, can run any application on the device, and can delete, modify, or insert any content in data stored in the device. The mobile devices are power-limited. When a device visits a location, it can find the location authority (or authorities) for the location. We also assume that the device can access the public key for the location authorities. The location authorities for a given location can check the presence of devices in that area. Devices can also check the presence of other devices, or the devices advertise their presence to neighboring devices. Communication between a device and the location authority or another device happens over wireless channels. At a later time, the user presents a location history claim to the auditor, providing her with a subsequence of her location provenance chain, containing an ordered subset of her previous locations and the times she visited these locations. The auditor uses the information stored in the location provenance chain to determine whether the claimed location history is supported by the proofs contained in the provenance chain.





\subsection{Threat Model}

We consider different classes of adversaries, and also combinations of these adversaries in a collusion attack. In our threat model, we will first discuss the assets, and then talk about the capability of attackers and the attacks they can launch.
 
\noindent\textbf{Assets.~}
The two main targets considered in our threat model are: the location proofs, and the chronological ordering of location proofs. An adversary should not be able to create a location proof for a location that the user has not visited. Also, even if the user has visited the location, an adversary should not be able to create a proof for a different (local) time than the actual time of visit. A false location proof is one that attests to the user's presence in a location not visited by the user, or the presence at a different time than the real time of visit. The order information is also critical and must be secured. An adversary should not be able to reorder the location proofs to change the sequence of location visits in the user's history. An additional target of malicious attackers can be the privacy of the user's location history. An attacker may want to create a dossier of users visiting a given location. For a given user, a malicious attacker may also want to learn the user's location history and the identities of other users it has encountered in the past.

\noindent\textbf{Attacker capabilities and attacks.~} Unlike previous work in the area \cite{luo2010hotmobile,saroiu2009hotmobile,waters2003secure}, we do not consider the location authorities as trustworthy. We assume that users, location authorities, or other witness devices present in the location and participating in the proof generation protocol can all be malicious. Users, locations, and witnesses can collude with one another. Typical attacks such as MAC address fingerprinting are prevented via known techniques such as MAC address cloning \cite{martinovic2007phishing}. A malicious user does not have access to the private key of a non-colluding user, witness, or location. We assume that well-known techniques are used to prevent denial of service attacks by adversaries who flood the location with proof requests and witnesses with endorsement requests. In addition, we assume that users do not have multiple identities and therefore no Sybil attacks occur \cite{sybil}.

We aim at making location provenance chains tamper-evident, as opposed to tamper-proof. Since the provenance information is carried at the User's mobile device, the user can always tamper with it. Without trusted hardware, it is impossible to prevent the user from tampering with provenance. Therefore, we focus on ensuring detection of different types of attacks on provenance. In the following, we list a number of attacks we want to prevent. 
\begin{itemize*}
\item[] \noindent\textbf{False presence.~} A malicious user can create a fake location proof without being physically present in the location. 

\item[] \noindent\textbf{False time (backdating, future dating, present dating).~} A location proof is generated for a user who has visited the location, but the proof's time is different from the time of visit. The backdating attack creates a proof for a past time, while the future dating attack, a proof is generated for a future time. In a post-dating attack, a location proof is planned to be issued at a future time (i.e., the location authority colludes with location to create a false proof with a future date, and premeditates to introduce the false proof at a later time).

\item[] \noindent\textbf{Reordering.~} The sequence of user location proofs is reordered to create a false path for the user.

\item[] \noindent\textbf{Implication.~} A location authority and/or witnesses falsely claim the user is/was present in the location

\item[] \noindent\textbf{False endorsement.~} A colluding witness endorses a user who is not present in the same location as the witness.

\item[] \noindent\textbf{Denial of presence.~} A user denies having visited a location.

\item[] \noindent\textbf{Proof switching.~} When a legitimate proof is used to create a false proof for a different location.

\item[] \noindent\textbf{Doppleganger.~} When the device is cloned or its keys are shared, to create a separate device that is capable of participating in protocols pretending as the device.
\end{itemize*}
	\section{A Secure Location Provenance Scheme}
\label{sec:scheme}

\noindent\textbf{Definitions.} To state that user $U$ has visited location $L$ at time $t$, the location authority prepares a location statement $LS$ as follows:
\begin{equation}
LS = < U, L, t >
\end{equation}
Depending on the secure localization scheme, the location statement can include further proof of the user's presence and interaction with the location \cite{luo2010hotmobile,saroiu2009hotmobile}.

The location authority creates a location proof $LP$ as follows: 
\begin{equation}
LP = < LS, s_{L} (LS)>
\end{equation}
where $S_{L}(LS)$ represents the digital signature computed on $LS$ using the location authority's private key.

An endorsed location proof, $ELP$ is defined as follows:
\begin{equation}
ELP = < LP, E >
\end{equation}
where $E$ is an endorsement defined as.
\begin{equation}
E = < ES, S_{w}(ES)>
\end{equation}
where $ES$ is an endorsement statement, and $S_{w}(ES)$ is a signature computed by the endorsing witness $w$ on $ES$.

The endorsement statement $ES$ is created by a witness $w$ to endorse $LP$. 
\begin{equation}
 ES = < w, U, L, t, h(LP), t_{e} >
\end{equation}
where $t_{e}$ is the signed endorsement timestamp given by the location and h is a cryptographic hash function. 

A location provenance entry, $LProv$ is defined as:
\begin{equation}
LProv = < ELP, C >
\end{equation}
where $C$ is an additional cryptographic construct used to provide chronological ordering of the location provenance entries.
 
Finally, a location provenance chain $LPC$ of size $n$ for user $U$ is defined as,
\begin{equation}
LPC = < LProv_{1}, LProv_{2}, \cdots, LProv_{n}>, 
\end{equation}
where for any $ 1\leqslant  i < j \leqslant n$, $U$ visited $L_{i}$ before visiting $L_{j}$.

\subsection{Location Provenance Protocol}
\begin{figure}[tbp]
\centering
\includegraphics[width=3.3in]{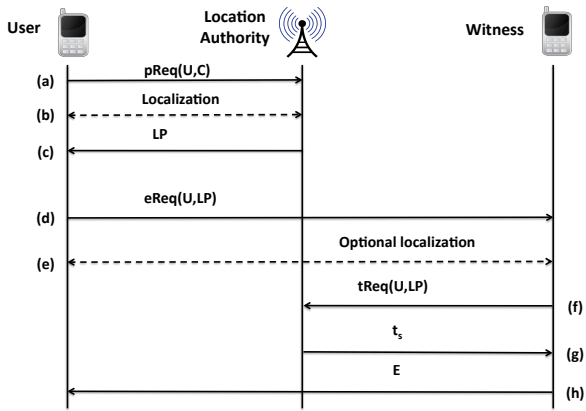}
\caption{Different steps of the Location Provenance Protocol.}
\label{fig:protocol}
\end{figure}

Creation of a new location provenance entry for a location the user visits takes place in the following steps, as illustrated in Figure~\ref{fig:protocol}:\\

\noindent\textbf{a. Location authority discovery and proof request.~}
When a user arrives at a location, she initiates the protocol for getting a location proof from the location authority associated with the location. Each location is identified by a unique global identifier. We assume that location authorities and their identifiers are publicly available (perhaps via a lookup), or location authorities periodically broadcast their information to nearby devices. The user obtains the identity of the location authority and sends a request for location proof ($pReq$) message to the location authority. This message includes the identity $U$ of the user represented by her public key \cite{saroiu2009hotmobile}, or by anonymized identifiers \cite{luo2010hotmobile}. It also includes chronological ordering information $C$ from the last entry of the user's location provenance chain. 

\noindent\textbf{b. Secure localization.~}
Upon receiving the $pReq$ message, the location authority runs a secure localization step to determine whether the device is actually present in the location. We do not require any specific localization scheme -- any secure scheme including distance-bounding or visual scanning can be used in this step. 

\noindent\textbf{c. Location proof generation.~}
After being satisfied that the user is present in the location, the location authority generates the location proof, consisting of the location statement and a signature attesting the statement. The statement includes the identities of the user and the location, and the local timestamp of the visit. The location authority also creates the new chronological ordering information for the new location provenance entry. Next, the location authority sends this back to the user.

\noindent\textbf{d. Witness discovery and Endorsement request.~}
The user next contacts a witness to endorse the location proof. To obtain an endorsement, the user sends the message $eReq$ to the witness. The devices willing to serve as a witness can advertise this to nearby devices. 

\noindent\textbf{e. Witness localization.~}The witness perform its own distance bounding or other secure localization to verify whether the user is co-located with it. 

\noindent\textbf{f. Endorsement message creation.~} After verifying user presence, the witness creates an endorsement message $ES$ and sends a request $tReq$ to the location authority for timestamping. 

\noindent\textbf{g. Endorsement timestamping.~}The location authority creates a signed timestamp $t_{s}$ and then returns it to the witness. It does not approve the timestamp if the timestamp request came a long time after the original proof was issued.

\noindent\textbf{h. Endorsement signature generation.~} If the endorsement timestamp $t_{e}$ is after $t$, but not by a great extent, the witness accepts the timestamp. Finally, the witness creates the signed endorsement $E$ and sends it to the user. The witness can use its private key to create the endorsement signature. If the witness wants to remain anonymous, she can also sign the endorsement with a group signature key \cite{chaum91group} rather than its private key. The user puts the endorsed location proof in the new location provenance entry for the location.\\

Additionally, the location maintains a publicly visible list where hashes of each of the location proofs are inserted. At every epoch, it publishes the current state of this list along with a signature. The purpose of this publicly available list of proof hashes is to prevent backdating and future-dating attacks (we discuss it further in Section \ref{sec:analysis}). 
 

\subsection{Private Location Proofs.~}
The position of a user can be described using multiple granularities. For example, a user's location can be described in terms of the city, neighborhood, block, street address, or the exact position. To different auditors, the user may want to reveal the her in different granularities. 
To allow users to prove their locations to an auditor only at the desired granularity, we introduce the notion of private location proofs. For this, we use two mechanisms -- blinded location statements and proxy proofs.

Suppose the user's location $L$ can be described in $n$ different granularities, i.e., $L= \{l_{1}, l_{2}, \cdots, l_{n}\}$. To allow the user to choose any one of them and prove that to the auditor, a naive solution is to create multiple location statements, each with a different granularity of the user's location, and then create separate proofs for each \cite{luo2010hotmobile}. However, it increases the burden on the location authority. Instead, we create a single proof where the location granularities are blinded as follows: the location authority includes the $n$ location attributes in the location statement $LS$, and also includes cryptographic commitments for them \cite{blum1981}. That is, for each granularity $l_{i}$, the location authority computes a commitment $comm_{i}$ as follows: $comm_{i} = h(l_{i}, r_{i})$, where $h$ is a one-way hash function, and $r_{i}$ is a random nonce. The private location statement $LS_{p}$, is constructed as follows: 
\begin{equation*}
LS_{p} = <U, L, t, (comm_{1}  | \cdots | comm_{n}) | (r_{1} | \cdots | r_{n})>
\end{equation*}
where $|$ is the concatenation operator. 
When computing the signature over the $LS_{p}$, the location authority does not use the actual location granularities $l_{i}$, but includes the commitments $comm_{i}$. This allows removal of one or more location attributes from the location statement when it is revealed to the auditor. The user provides the location granularity $l_{i}$ and the corresponding random nonce $r_{i}$. Using these, the auditor can verify the unblinded location entry matches the corresponding commitment that was used to create the location statement.  



This, however, still does not prevent a curious auditor from guessing the user's location based on the position of the location authority. To prevent that, we propose using proxy proofs that are issued by a third party location authority. Suppose that the location authority in a city block is requested to provide a private proof with granularities ranging from the city to the exact location of the user. The location authority for the block can contact the location authority for the city for help with this. For this to work, the two location authorities must trust each other. Upon creating a location proof, the block's location authority sends it to the city location authority. The latter verifies the proof, and then re-signs it, replacing also the identity of the block location authority in the location statement. The new location proof is sent to the block location authority who sends it to the user. On receiving such a proof, an auditor cannot guess which city block the user has been, unless the user reveals it to the auditor.

\subsection{Creating Location Provenance Chains}

A location provenance chain consists of location provenance entries and chronological meta data that shows the order of these entries. As discussed earlier, users should be able to reveal any arbitrary subsequence of their past locations and prove their chronological ordering. For this, we propose two schemes -- a signed hash chain scheme, and a Bloom filter based scheme.

\subsubsection{Hash Chain-based Ordering}
In a signed hash chain scheme, each of the provenance entries contain chronological ordering metadata linking the entry directly with the previous entry. For example, we can store a signed hash value at each location provenance entry, which is computed by taking the hash value from the previous location, concatenate it with the hash of the current location provenance entry, and then sign it. The one-way nature of hash functions makes it impossible to modify the ordering information. Each hash is signed by the corresponding location authority, and can be verified by the auditor. To create a new location provenance entry $LProv_{i}$, the location authority requests the chronological ordering metadata $C_{i-1}$ from the last location provenance entry, and then calculates the new metadata $C_{i}$ using the following equation:
$C_{i} = s_{L} ( h(LP_{i}) | C_{i-1})$, where $s_{L}$ indicates a digital signature using the location's private key.

When presenting the chosen subsequence to the auditor, the user removes the location proofs corresponding to locations she wants to keep private. However, the signed hash $C$ values for all fields are retained, as well as the corresponding $h(LP)$ hashes. Utilizing the one way nature of hash chains, the auditor can verify the order of the locations by recreating and validating the $C$ values. The downside of this scheme is that, at the worst case when the head and tail locations in the list are revealed, the auditor has to traverse the entire chain regardless of how many elements from the chain have been revealed by the user.


\subsubsection{Accumulator-based Ordering}
Instead of linking two provenance entries using hashes, we introduce a novel accumulator based scheme to prove chronological ordering. The key idea is to accumulate the hash values of all location proofs obtained so far in the accumulator for the current provenance entry. An accumulator in this context is a construct that can be used for set operations. To create a new location provenance entry, the location authority requests the last value of the accumulator from the user. It then hashes the current location proof, inserts it into the accumulator, and signs it, and sends it back to the user. Two accumulators can be compared as follows: the accumulator for later entries are supersets of that of previous entries.
 
For example, suppose the user visited locations $L_{1}, L_{2}, L_{3}, L_{4}$, in that order. The accumulator $C_{1}$ stored in $LProv_{1}$ has the hash of $LP_{1}$. To create $C_{2}$ for $LProv_{2}$, the location authority takes $C_{1}$, and inserts the hash of $LP_{2}$ in it, to obtain $C_{2}$. Similarly, $C_{4}$ has the hashes of the location proofs $LP_{1}, LP_{2}, LP_{3}, LP_{4}$. Now, given any two accumulators $C_{i}$, and $C_{j}$, we can prove their order by checking which one of them is the subset of the other. For example, $C_{1}$ is a subset of $C_{2}$, $C_{3}$, and $C_{4}$. Also, $C_{2}$ is a subset of $C_{3}$ and $C_{4}$, but not of $C_{1}$. In other words, given two accumulators $C_{i}$ and $C_{j}$, the auditor can verify that $L_{i}$ was visited prior to visiting $L_{j}$ if and only if $C_{i}$ is a subset of $C_{j}$ Thus, without having any explicit chaining between the entries, we can check the ordering of the the location proofs. Unlike hash chains, accumulators have the nice property that we no longer have to keep the entire provenance chain to show order of any arbitrary subsequence. Rather, a user can reveal only the elements in the subsequence and the auditor can verify their order by checking the accumulators. 

While any accumulator scheme that allows set membership checks will work here, we propose using a Bloom-filter based accumulator. A Bloom-filter is a probabilistic data structure with no false negatives, which can be used to test set membership in constant time \cite{bloom70filter}. In this construction, we keep a single Bloom filter per location provenance entry. Using a Bloom filter has certain advantages. Since a Bloom filter is essentially a bit array, we can verify order quickly, by checking which bit array is a subset of the other, using fast bitwise operations (i.e., A is a subset of B, iff A AND B = A). Inserting a new location proof and membership check can be done in constant time. The downside of using a Bloom filter is the false positive rate. However, we can reduce that by allotting more bits to the Bloom filter. Also, given two Bloom filters, an adversary can compute the number of bits in which they differ, and thereby calculate number of locations visited between these two locations. Depending on the scenario, this may or may not be a problem. We argue that the Bloom filter based scheme, while incurring more space than the hash chain based accumulator, is preferable when revealing smaller fractions of visited locations. This is because with Bloom filters, only the filters from the revealed provenance entries are needed, while in the hash chain, the entire chain is needed in the worst case.

\subsection{Audits}
We assume that, the auditor has access to the public keys of the location authorities and the group keys of different groups. To initiate an audit, the user presents the auditor with a location history claim and the supporting location provenance information. The location history claim consists of an ordered sequence of location claims. Each location claim consists of the identifier for a location and the time of visit. 

The auditor's task is two-fold: the auditor first checks if each location claim matches the corresponding location provenance entry. To verify this, the auditor checks the integrity of the location provenance entry, verifies the signatures, matches the endorsements with the location proofs, and finally checks whether the location and time claimed by the user matches the location and time mentioned in the location proof(s). To prevent back or future-dating attacks, the auditor also checks if a location proof is in the published list of proof hashes of the corresponding epoch for that location. 
Next, the auditor verifies whether the claimed order of the locations matches the given location provenance entries. To do that, the auditor uses the chronological ordering information given in the location provenance entries. 
	\section{Security Analysis}
\label{sec:analysis}

In this section, we present an analysis of the security properties of our schemes. We start by enumerating the different types of attackers and collusion among them. Then we analyze how our schemes can protect against such attacks.

\noindent\textbf{Symbols.~} We define the following symbols: honest user: $U$,  malicious user: $\bar{U}$, honest location: $L$, malicious Location: $\bar{L}$, honest witness: $W$, malicious Witness: $\bar{W}$ Using the above symbols, we can have the following three single party attacks: Malicious user ($\bar{U}LW$), malicious location ($U\bar{L}W$), and malicious witness ($UL\bar{W}$). We can also define three two party collusions, and one three party collusion. Table \ref{table:collusionmodel} summarizes the different attack scenarios and the corresponding threats.

\begin{table*}[tbp]
\centering
\begin{tabular}{|c|c|p{3in}|}
\hline \textbf{Model} & \textbf{Description} & \textbf{Threat/Attack} \\ 
\hline $ULW$ & Everyone is honest & No collusion. \\ 
\hline $\mathbf{\bar{U}}LW$ & Malicious user & False location proofs, reordering, doppleganger, denial or presence, proof switching.\\ 
\hline $U\mathbf{\bar{L}}W$ & Malicious location authority & Denial of service, implication. \\ 
\hline $UL\mathbf{\bar{W}}$ & Malicious witness & False endorsement, privacy. \\ 
\hline $\mathbf{\bar{U}\bar{L}}W$ & user and Location collude & False location proofs.\\ 
\hline $\mathbf{\bar{U}}L\mathbf{\bar{W}}$ & user and Witness collude & False endorsement.\\ 
\hline $U\mathbf{\bar{L}\bar{W}}$ & Location and Witness collude & Implication.\\ 
\hline $\mathbf{\bar{U}\bar{L}\bar{W}}$ & Everyone collude & False proofs.\\ 
\hline 
\end{tabular} 
\caption{Threat model}
\label{table:collusionmodel}
\end{table*}

\noindent\textbf{Single entity attacks.~} The three single entity attacks are $\mathbf{\bar{U}}LW$, $U\mathbf{\bar{L}}W$, and $UL\mathbf{\bar{W}}$. In $\mathbf{\bar{U}}LW$, a malicious user $\mathbf{\bar{U}}$ creates false location proofs. But if $L$ and $W$ are honest, this attack does not succeed, since an honest $L$ will not sign a location proof if it does not detect the user in its proximity.  Also, an honest $W$ will not endorse a location statement not accompanied by a proof from the location. $U\mathbf{\bar{L}}W$ attacks also fail since the dishonest location authority $\mathbf{\bar{L}}$ cannot create a false proof when the user is not present in the location. This is because the secure localization step involves interaction between the $U$ and $\mathbf{\bar{L}}$, and $\mathbf{\bar{L}}$ cannot falsify the interaction without knowing $U$'s keys \cite{waters2003secure}.  An honest witness will also not endorse a proof unless it can detect $U$'s presence. The only thing $\mathbf{\bar{L}}$  can do is to provide a false timestamp. However, an honest witness will not endorse a proof if the timestamp differs a lot from its own location proof timestamp. Finally, a malicious witness ($UL\mathbf{\bar{W}}$) cannot do any harm other than denial of service.

\noindent\textbf{Multiple entity collusion.~} Two entity attacks involve scenarios where any two of the entities in user, location, and witness collude together. There are three possibilities: $\mathbf{\bar{U}}\mathbf{\bar{L}}W$, $\mathbf{\bar{U}}L\mathbf{\bar{W}}$, and $U\mathbf{\bar{L}}\mathbf{\bar{W}}$. 

The user and location can collude ($\mathbf{\bar{U}}\mathbf{\bar{L}}W$) to create false location proofs. First, they can create a proof offline even when the user is not present at the location. Such a proof will not be endorsed by an honest witness, and hence can be detected by an auditor. Next, the user may be present in the location, but the location proof can be given a backdated or future-dated timestamp by the location.  There are several ways of detecting and preventing such attacks. We can prevent backdating by requiring each location to publish a list of users that has visited the location. While a simple list would serve the purpose, that violates user privacy. So, the list should be hard to decode without knowing the user (i.e., given just the list, it is difficult to figure out the identities of the users. But given a location proof, it can be checked whether the location proof was issued by the location at the given epoch).  Locations can divide time into epochs. At the end of each epoch, the location publishes the list of location proofs it issued (or rather, the hash of the location proofs it issued.). Given just the hashes, it is difficult to figure out the location proofs due to the one-way nature of hash functions. However, if one is presented with a location proof, computing the hash and checking if it is included in the list is not very difficult. The location can actually publish a signed accumulator of all hashes of the location proofs for a given epoch. This can prevent backdating attacks in the following manner: suppose that the location and the user collude to create a backdated location proof at time $t'_{u}$ that corresponds to epoch e. However, the location has already published the accumulator for that epoch, so the auditor can go and check the accumulator for that epoch using the location proof. Also, witnesses can check to see if the location proof timestamp is different from its own location proof timestamp. The same applies to future dating. If the location gives a user a future timestamp, it must also give all the witnesses a future timestamp to evade detection. 

A malicious user and a colluding witness ($\mathbf{\bar{U}}L\mathbf{\bar{W}}$) cannot create false location proofs if the location authority is honest. Similarly, a malicious location authority can collude with a dishonest witness ($U\mathbf{\bar{L}}\mathbf{\bar{W}}$) and attempt to implicate an honest user. However, if the user never participated in a proof protocol with the location authority, such an attack will not work. Finally, the location authority can give a user a backdated or a future dated timestamp. A colluding witness can endorse such a false timestamped proof. However, the user chooses its own witness, and to effectively launch the attack, a location authority will have to collude with a majority of the users in the area. The location proof timestamp will also not will also not match with the timestamp of the published location proof accumulator for the epoch.

Finally, all three entities can collude ($\mathbf{\bar{U}\bar{L}\bar{W}}$) to generate false location proofs and endorsements. Even then, a backdated attack can be foiled if the auditor checks the published accumulator for the epoch corresponding to the proof timestamp. A future-dated timestamp will also not match with the epoch timestamp. The only attack that is possible here is a post-dating attack, i.e., when the user, location authority, and witness collude to create a location proof for a future time, but does not publish it in the epoch report. After creating the proof, the location caches it, and later includes it in the future epoch's accumulator, at a time when the user is actually at a different location. Such attacks are hard to prevent. An auditor can require a stricter proof model involving statements from more than one location authorities. 

\noindent\textbf{Other attacks.~} In a doppelganger attack, the user shares her secrets with an accomplice who uses them to impersonate the user and obtain a location proof from a location where the user is not present. For example, the user can share her keys with an accomplice, who can use them to get proofs on behalf of the user. This can be prevented via radiometric signatures, which are unique \cite{brik2008wireless}.


	\section{Evaluation}
\label{sec:evaluation}

To evaluate the performance and feasibility of our location provenance scheme, we created a proof of concept system prototype. We implemented the location provenance scheme in Java 1.5. We developed a location provenance application on the Android 2.2 platform and deployed the application on Motorola Droid 9 mobile phones. Each phone had a ARM Cortex A8 processor with 550 mHz clock speed. The application also supported the witness mode, where the mobile phone acted as the witness for another phone. We also implemented a location proof generation application for the location authority. This was also on the Android 2.2 platform, and was tested on the same Droid phones. Note that we decided to evaluate the performance on mobile phones so that location authorities can be deployed easily and cheaply. For cryptographic operations, we used 1024-bit DSA signatures and 160-bit SHA-1 hashes from the Bouncy Castle cryptographic library \cite{bouncycastle}.

\begin{figure}[tbp]
  \centering
\includegraphics[width=3.2in]{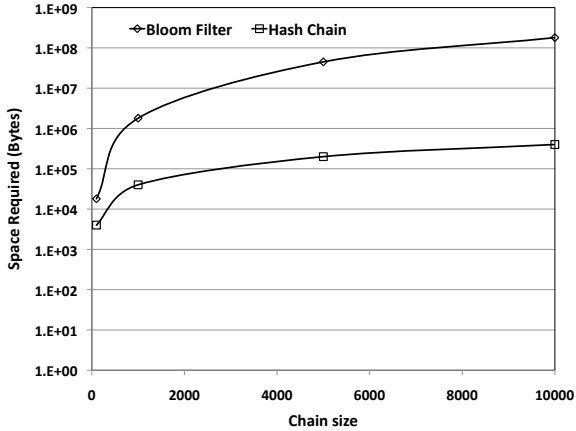}
  \caption{Comparison of space requirements of Bloom Filter and Hash Chain schemes.}
 \label{fig:spacereq}  
\end{figure}

\subsection{Experiments}
\noindent\textbf{Goals.~} The goal of our experiments was to evaluate the feasibility of deploying our schemes on existing off-the-shelf mobile devices and to compare the Bloom filter based scheme and the hash chain based scheme for securing the chronology of proofs. For each scheme, we measured the space requirements and overhead of location provenance and private location proofs. We also measured the proof generation times and audit performance under different configurations and location granularities.

\noindent\textbf{Space requirement.~} Figure \ref{fig:spacereq} shows the space requirements for the provenance chain ordering mechanism for the Bloom Filter scheme. For the Bloom Filter, we used a 0.1\% false positive rate. The hash chain based scheme requires 40 bytes per entry (the size of a DSA signature), regardless of the size of the provenance chain. The Bloom filter based scheme, however, requires more space, depending on the total size of the chain. For example, for a chain with 1000 locations, we need 40 bytes per location provenance entry in the hash chain scheme, but to have a 0.1\% false positive rate, we need 1797 bytes for the Bloom filter stored with each entry. 
\begin{figure}[tbp]
  \centering
\includegraphics[width=3.2in]{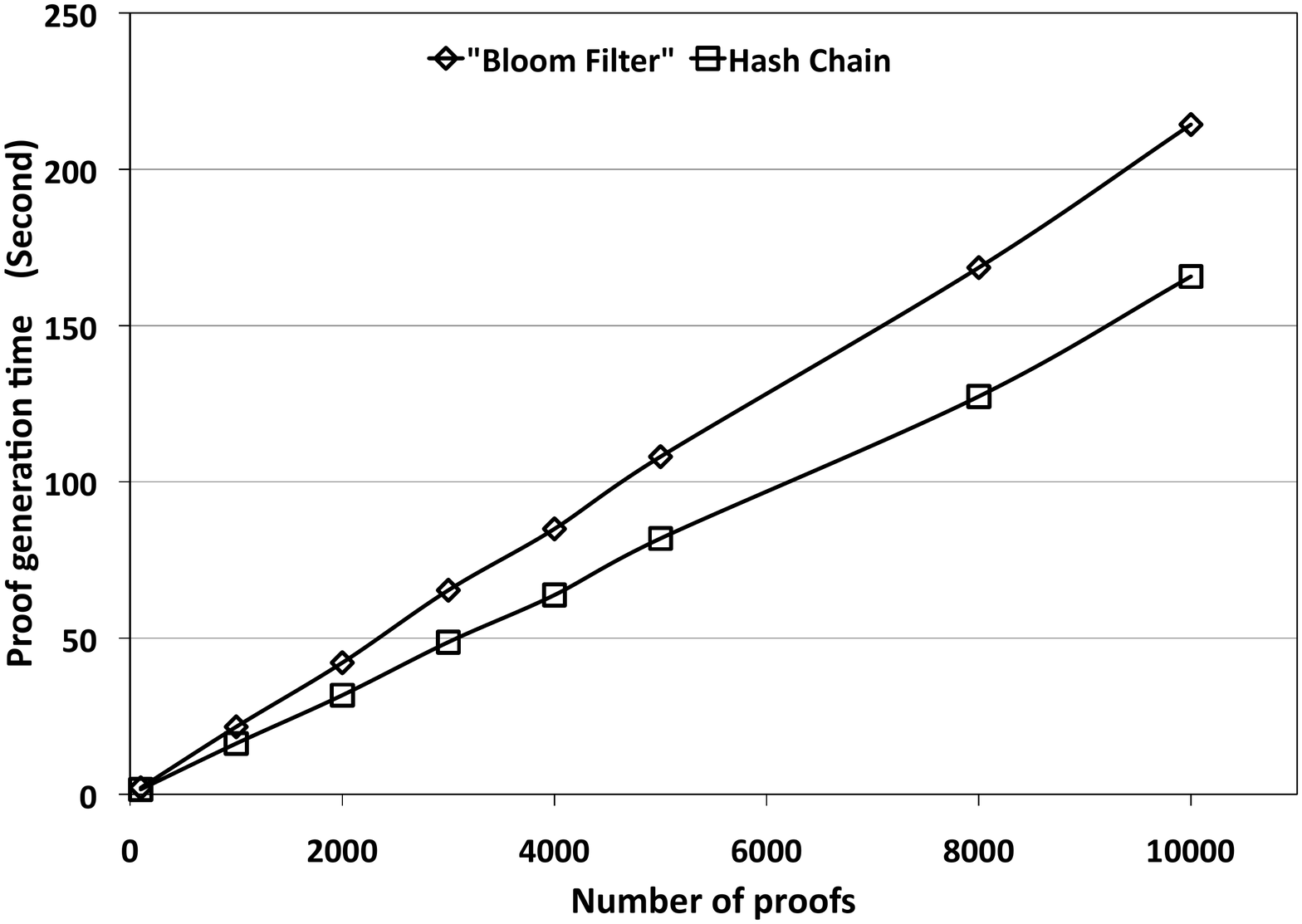}
  \caption{Time to generate location proofs. For the Bloom filter scheme, generation rate is 46 proof/second, while with hash chains, the rate is 60 proofs/second.}
 \label{fig:protocoltime}  
\end{figure}

\begin{figure}[tbp]
  \centering
\includegraphics[width=3.23in]{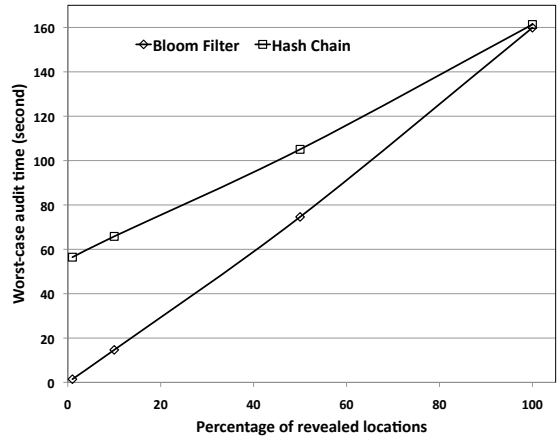}
  \caption{Worst case audit time for a 10,000-entry location provenance chain, when different percentages (1\%, 10\%, 50\%, and 100\%) of the locations are revealed.}
 \label{fig:audit1000chain}  
\end{figure}

\noindent\textbf{Proof generation time.~} Figure \ref{fig:protocoltime} shows the time to generate location proofs by the location authority. The hash-chain based scheme is slightly faster, and we obtained proof generation rate of 60 proofs/second on the Droid phones. With the Bloom filter based scheme, the proof generation rate is 46 proofs/second. Since we achieved this rate even on low-end phones, it shows that our scheme is feasible. In practice, larger locations will have powerful and dedicated computers that can generate proofs even faster.

\begin{figure*}[ht]
\centering
\subfigure[1\% locations revealed]{
\includegraphics[width=3.1in]{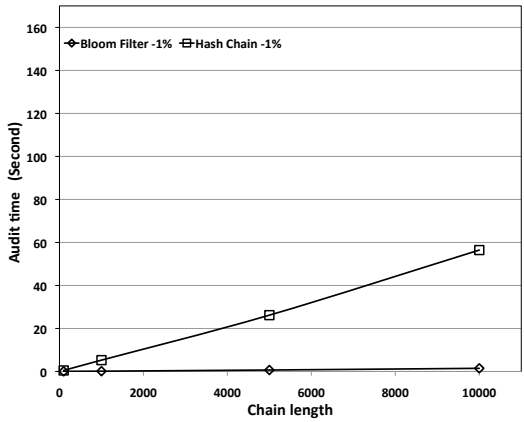}
\label{fig:audit1}
}
\subfigure[10\% locations revealed]{
\includegraphics[width=3.1in]{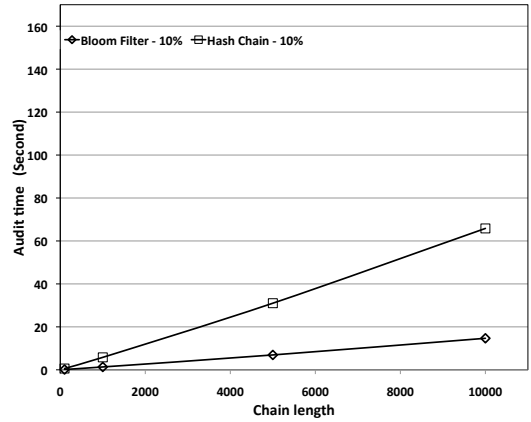}
\label{fig:audit10}
}
\subfigure[50\% locations revealed]{
\includegraphics[width=3.1in]{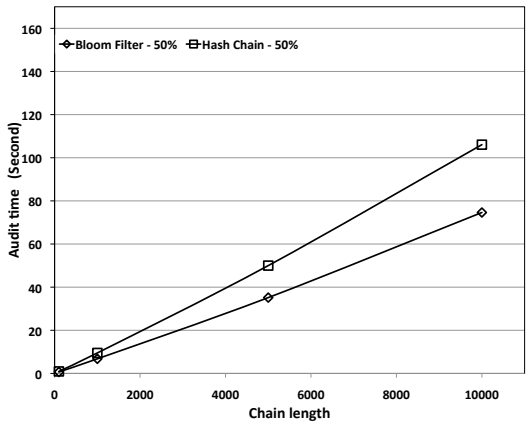}
\label{fig:audit50}
}
\subfigure[100\% locations revealed]{
\includegraphics[width=3.1in]{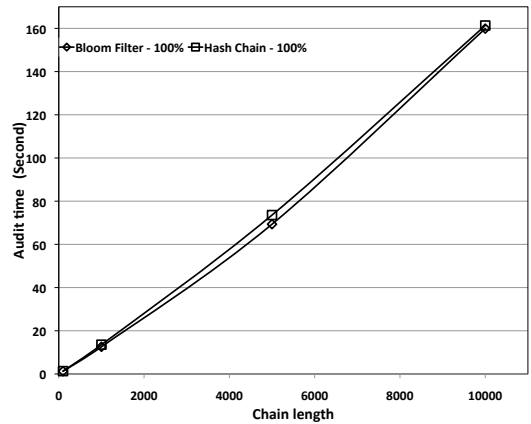}
\label{fig:audit100}
}
\caption[Optional caption for list of figures]{Comparison of worst-case audit times for the Bloom-filter and Hash-chain schemes, when 1\%, 10\%, 50\%, and 100\% of location entries are revealed to the auditor. When smaller fraction of locations are revealed, the Bloom filter scheme outperforms the hash chain scheme.}
\label{fig:AuditPerformance}
\end{figure*}

\noindent\textbf{Overhead of private location proofs.~} For private location proofs, additional space required for each granularity in the location statement includes a SHA-1 hash for the commitment (20 bytes), and a random nonce (4 bytes), resulting in a total of 24 bytes additional space per granularity. It also slightly increased proof generation time by approximately 0.4 milliseconds per granularity (approx. 1.8\% overhead for Bloom filter and 2.4\% overhead for the hash chain scheme).
	
\noindent\textbf{Audit performance.~} We next measured the worst case audit performance of the hash chain and Bloom filter based provenance schemes. We created an Audit application to verify the integrity and consistency of the location proofs as well as location provenance. The application was tested on the Droid phones. For both schemes, we measured the audit performance for different sized chains, with different number of items revealed to the auditor. Figure \ref{fig:audit1000chain} shows the relative performance of the two schemes for a 10,000 location chain, when 1\%, 10\%, 50\%, and 100\% of the locations contained in the provenance chain are revealed to the auditor. We see that when all the locations are revealed, both schemes perform on an equal basis, requiring about 15 milliseconds per location proof. However, when a small number of location proofs are revealed to the auditor, the Bloom filter based scheme outperforms the hash chain scheme. For example, when 1\% locations are revealed, the Bloom filter based scheme still requires 15 milliseconds per proof, but the hash chain based scheme requires 564 milliseconds. This is because, with the Bloom filters, the auditor only needs to check the revealed location proofs to verify the order. But with the hash chain scheme, the auditor, at the worst case (when locations near the beginning and end are revealed), needs to traverse the entire hash chain to verify the ordering. We further illustrate this in Figure.~\ref{fig:audit1}--\ref{fig:audit100}, where we show the comparison of audit times when 1\%, 10\%, 50\%, and 100\% locations are revealed, respectively. In all but the last case, the Bloom filter based scheme outperforms the hash chain scheme.

	\section{Related Work}
\label{sec:related}
Secure location verification and location privacy have been explored by many researchers. Gonzalez-Tablas \etal provide an overview of research on secure location verification \cite{gonzalez2008pervasive}. Research on secure location verification can broadly be divided into two categories: verifying current location (secure localization), and verifying past location (location proofs). In the first category, the goal is to securely determine the current location of the user, while in the second category, the focus is on verifying the past locations of a user using location proofs.

\subsection{Secure Localization}
Gabber \etal described early work on proving location of devices \cite{gabber1998ccs}. The focus of this work was to verify the location of customer equipment (such as satellite receivers). The three schemes described here used telephone caller ID, GPS, and cellular tower signals respectively. Unfortunately, all of these methods can be bypassed by malicious attackers \cite{luo2010hotmobile,saroiu2009hotmobile}. Denning \etal presented a location verification scheme using location signature sensors (LSS). This requires specialized hardware that can obtain signal from a large number of GPS satellites and combine them to create a signature. Verifiers also concurrently obtain a signature from the satellites, and use that to determine the device's location \cite{denning1998location}. However, this scheme does not provide any strong binding between user identities and the LSS signature, which in turn can allow spoofing and impersonation attacks \cite{luo2010hotmobile}.

Brands \etal introduced distance bounding protocols for determining proximity of a device to a location \cite{brands93eurocrypt}. This scheme uses the round-trip delay of the challenge response protocol between the user and the location authority. Using distance bounding, a location authority can check the minimum distance between it and the user. However, this scheme is prone to spoofing since it does not bind the user's identity to the proof \cite{brands93eurocrypt}. Sastry \etal introduced a secure location verification scheme that used a combination of radio frequency signal and ultrasound to determine the location of the user \cite{sastry2003secure}. However, this is subject to wormhole attacks \cite{luo2010hotmobile}. Researchers have since then extensively studied distance bounding as a practical and efficient scheme for secure localization  \cite{rasmussen2010distance}. Vora \etal discussed location verification using radio broadcast in wireless sensor networks \cite{vora2006radio}. Distance bounding and other secure localization techniques can be effectively used in the secure localization phase of our scheme.

\subsection{Location Proof}
The notion of secure unforgeable location proofs was discussed by Waters \etal \cite{waters2003secure}, who discuss a secure scheme which a device can use to get a location proof from a location manager. However, it requires users to know the auditors \textit{a prior}. Lenders \etal presented a secure geotagging service for verifying the location and timestamp for user-generated content \cite{lenders2008hotmobile}. Users wishing to timestamp their content send a hash to a location-timestamp authority, which then determines the user's location and then issues a location-time signature using the hash. Here, the focus is on certifying the user supplied content rather than the user's location. The signature does not bind the content to the user's identity. Their system is also highly centralized.

 Saroiu \etal\cite{saroiu2009hotmobile} proposed a mechanism for creating secure location proofs. In this system, users and wireless access points exchange their signed public keys to create a timestamped proof of visit. Saroiu \etal \cite{saroiu2010hotmobile} and Gilbert \etal \cite{gilbert2010hotmobile} described trustworthy sensors where the reading from a sensor is attested using hardware based TPM or software based virtual machines. Luo \etal described a scheme for creating privacy-preserving location proofs \cite{luo2010hotmobile}. Instead of using public keys as identity, the user commits to a random nonce when initiating the location proof protocol with the proof provider. They support multiple location granularities, but the location information for different granularities are separately encrypted and decryption keys have to be transmitted to the user. They also described six design goals for flexible location proof systems : scalability, application-agnostic proofs, proactive collection of proofs, user anonymity and privacy, and the ability to run on regular hardware. Our scheme for location provenance meet all these goals. However, our work differs from most of the existing schemes for two main reasons: first, most of the existing research do not consider that the location authority to be malicious. The schemes also do not protect against collusion attacks. In contrast, our witness-endorsement location provenance scheme is resistant against collusion between locations. Second, many existing schemes do not provide mechanisms for verifying the order of location proofs. For schemes which do provide ordering (such as Path-stamps \cite{gonzalez2003pathstamps}), there is no support for revealing only a subset of locations to auditors, which we do provide using our location provenance scheme.

\subsection{Location History}
Ananthanarayanan \etal discussed a framework for collecting and storing sequence of user locations \cite{ananth2009startrack}. In their StarTrack system, sequence of a user's  location and time entries are stored in tracks which can later be shared compared, clustered, and queried. While tracks are similar to location provenance chains, security issues are not considered, making tracks vulnerable to attacks by malicious users. Zugenmaier \etal introduced the notion of location stamps \cite{zugenmaier2001enhancing} for cell phones. The stamps provide proof about the location of the user at a certain time. Based on this idea, Gonzalez-Tablas \etal developed the notion of Path-stamps \cite{gonzalez2003pathstamps}, where the location history of the user is considered. Here, a sequence of location stamps, i.e., location proofs, are combined by creating a hash chain. A big difference with our work is that they considered the proof issuer to be trustworthy and not collude with the users. Their system is dependent on a centralized proof issuer, which makes it non-scalable and inefficient. The path stamps protocol also requires each user to possess a specialized hardware that is used to authenticate the user. This requirement makes the system difficult and expensive to deploy. Path stamps protocol does not support publication of partial path, therefore users must reveal their entire path to auditors even when they are proving a subset of their path. In contrast, our schemes allows users to have privacy by enabling them to prove any arbitrary subset of their location history.  Finally, Interaction-based missed connection services have been described by Manweiler \etal \cite{manweiler2009smile}. Here, two mutual strangers can use the SMILE protocol to establish shared knowledge, which can later be used to prove that they have met before. Such a scheme can be used to make our endorsement schemes truly anonymous.

\subsection{Secure Provenance in Other Domains}
Provenance has been studied by researchers to provide information about the lineage, origin, and transformation of data objects. Simmhan \etal provided a survey of the use of data provenance in scientific computing \cite{simmhan05}. Most of the systems discussed there were scientific computing systems, where security issues are not handled.  Muniswamy-Reddy described a provenance aware file system named PASS \cite{muniswamy06usenix}. It allowed automated recording of data provenance for files. Hasan \etal discussed secure provenance for files \cite{hasan2007sss}, and developed an efficient system for recording file system provenance in a secure manner \cite{hasan2009preventing}. Different aspects of secure data provenance have been discussed by other researchers \cite{lu2010secure, syalim2010preserving}. While conceptually similar to secure data provenance, secure location provenance introduces new challenges and attack vectors, that are not a concern in data provenance. For example, in secure location provenance, we need to ensure the user's physical presence in the location. This is not a concern in data provenance. However, with modifications, the location provenance scheme described in this paper can be effectively applied to tracking data movement in cloud computing. 
    \section{Conclusion}
\label{sec:conclusion}

Secure location provenance allows verification of location history, enabling trustworthy location-aware services. The ability to verify location history claims can lead to new applications in many domains. In this paper, we analyzed the secure location provenance problem and introduced techniques for making location proofs resistant against collusion. Our location proof scheme allows a user to prove her location in different granularities to different auditors. To protect against forgery of location chronology, we presented two schemes based on hash chains and Bloom filters. Through experimental results from our proof of concept implementation on the Android-based Droid phones, we demonstrated that such schemes are practical on today's mobile devices. Our future work includes the developing a theoretical framework for the different degrees of trust for location proofs.

\section*{Acknowledgements}
This material is based upon work supported by the National Science Foundation under Grant \#0937060 to the Computing Research Association for the CIFellows Project. 
\bibliographystyle{abbrv}
\bibliography{locprov,provenance}

\end{document}